\documentclass[english,pra,superscriptaddress,final,lengthcheck]{revtex4-1}
\usepackage[T1]{fontenc}
\usepackage[latin9]{inputenc}
\usepackage{color}
\usepackage{babel}
\usepackage{textcomp}
\usepackage{amsmath}
\usepackage{amssymb}
\usepackage{graphicx}
\usepackage[unicode=true,pdfusetitle,
 bookmarks=true,bookmarksnumbered=false,bookmarksopen=false,
 breaklinks=false,pdfborder={0 0 1},backref=false,colorlinks=true]
 {hyperref}

\makeatletter
 \@ifundefined{textcolor}{}
 {%
   \definecolor{BLACK}{gray}{0}
   \definecolor{WHITE}{gray}{1}
   \definecolor{RED}{rgb}{1,0,0}
   \definecolor{GREEN}{rgb}{0,1,0}
   \definecolor{BLUE}{rgb}{0,0,1}
   \definecolor{CYAN}{cmyk}{1,0,0,0}
   \definecolor{MAGENTA}{cmyk}{0,1,0,0}
   \definecolor{YELLOW}{cmyk}{0,0,1,0}
 }

\makeatother

\begin{document}

\title{Quantum simulations of localization effects with dipolar interactions}

\author{Gonzalo A. \'Alvarez}

\email{gonzalo.a.alvarez@weizmann.ac.il}

\selectlanguage{english}%

\affiliation{Department of Chemical Physics, Weizmann Institute of Science, Rehovot,
76100, Israel}

\affiliation{Fakult\"at Physik, Technische Universit\"at Dortmund, D-44221,
Dortmund, Germany}

\author{Robin Kaiser}

\email{robin.kaiser@inln.cnrs.fr}

\selectlanguage{english}%

\affiliation{Universite de Nice Sophia Antipolis, CNRS, Institut Non-Lineaire
de Nice, France}

\author{Dieter Suter}

\email{dieter.suter@tu-dortmund.de}

\selectlanguage{english}%

\affiliation{Fakult\"at Physik, Technische Universit\"at Dortmund, D-44221,
Dortmund, Germany}

\date{\today}
\begin{abstract}
Quantum information processing often uses systems with dipolar interactions.
We use a nuclear spin-based quantum simulator, to study the spreading
of information in such a dipolar-coupled system and how perturbations
to the dipolar couplings limit the spreading, leading to localization.
In {[}Phys. Rev. Lett. \textbf{104}, 230403 (2010){]}, we found that
the system reaches a dynamic equilibrium size, which decreases with
the square of the perturbation strength. Here, we study the impact
of a disordered Hamiltonian with dipolar $1/r^{3}$ interactions.
We show that the expansion of the coherence length of the cluster
size of the spins becomes frozen in the presence of large disorder,
reminiscent of Anderson localization of non-interacting waves in a
disordered potential. 
\end{abstract}
\maketitle

\section{Introduction}

The control of quantum mechanical systems is continuously gaining
interest in recent years \cite{Ladd2010,DiVincenzo2011,Simon2011,Aspuru-Guzik2012,Blatt2012,Bloch2012,Houck2012,Schulz2012},
mainly triggered by the pursuit of quantum information processing
and quantum simulations, which both have the potential of solving
computational problems more efficiently than classical computers \cite{Shor1994,DiVincenzo1995,Nielsen00}.
The realization of this potential requires precise control of large
quantum systems \cite{Hauke2012}. Controlling small quantum systems
has been explored extensively over the last years \cite{Ladd2010,DiVincenzo2011,Monz2011,Simon2011,Aspuru-Guzik2012,Blatt2012,Bloch2012,Houck2012,Schulz2012},
but control of large quantum systems is still very challenging. The
simulation of large quantum systems on classical computers is limited
to about 20 qubits if only pure states are considered \cite{Raedt04,zhang_modelling_2007}.
The typical classical algorithms can be also extended to calculate
the dynamics of mixed states if the initial state or the observables
are localized in a region smaller than the complete system. This approach
uses quantum parallelism of a single pure state evolution \cite{popescu_entanglement_2006,alvarez_quantum_2008}.
Present quantum technologies do not allow complete control of large
quantum states. So far large quantum systems were only addressed by
using ensemble quantum simulations with nuclear magnetic resonance
(NMR) experiments on solid state systems \cite{Suter04,Suter06,Lovric2007}. 

The main difficulties for controlling large quantum systems are the
lack of individual addressing of qubits, with important efforts in
progress with samples of ultracold atoms \cite{Bakr2009,Endres2011}.
For large systems decoherence is known to degrade the information
contained in the quantum state \cite{Zurek03}. Its rate increases
with the size of the quantum system, making the largest systems the
most susceptible to perturbations \cite{PhysicaA,JalPas01,Suter04,Suter06,Cory06,Lovric2007,sanchez_time_2007,Monz2011,Zwick2012}.
While these effects are known to affect the survival time of quantum
information, they also affect the distance over which quantum states
can be transmitted \cite{Pomeransky2004,Chiara2005,Keating2007,Burrell2007,Apollaro2007,Allcock2009,alvarez_nmr_2010,alvarez_localization_2011,alvarez_decoherence_2010,Zwick2011a,Zwick2012}.
Imperfections or disorder of the spin-spin couplings that drive the
state transfer can induce localization of the quantum information
\cite{Pomeransky2004,Burrell2007,Keating2007,Allcock2009} in a process
related to Anderson localization \cite{Anderson1958,anderson_local_1978}.
Whereas disorder induced inhibition of transport of non interacting
waves has been studied in various physical systems \cite{Hu2008,Chabe2008,Kondov2011,Jendrzejewski2012,Sperling2013},
the role of dipolar interactions is under theoretical investigation
\cite{Aleiner2011}. Here we study a 3D spin-network and demonstrated
experimentally a similar behavior by studying the localization effects
induced by the finite precision of quantum gate operations used for
transferring quantum states \cite{alvarez_nmr_2010,alvarez_localization_2011}. 

Reducing decoherence is a main step towards implementing large scale
quantum computers. Several techniques have been proposed for this
purpose, including dynamical decoupling \cite{5916}, decoherence-free
subspaces \cite{3045}, and quantum error correction \cite{3921,6581}.
These methods perform very well for small quantum systems \cite{6013,monz:200503,biercuk_optimized_2009,Du2009,Alvarez2010c,Souza2011,Souza2011a},
but they can be very challenging to implement in large quantum systems.
However, based only on global manipulations of spins, some of these
methods were successfully implemented in large quantum systems with
thousands of qubits \cite{Suter06,Lovric2007,sanchez_time_2007}.
The decoherence times were extended by almost two orders of magnitude.
Therefore, understanding the decoherence effects and their sources
on large quantum systems would help to optimize the control techniques
for fighting decoherence.

In this paper, we focus on understanding the impact of perturbations
with dipolar disorder on large quantum systems by quantum simulations
with solid state nuclear spin systems. These interactions depend on
the distance $r$ between the spins as $1/r^{3}$. In particular we
study the length scale of localization induced by perturbing the Hamiltonian
that drives the spreading of the information. Based on our previous
results and methods developed in Refs. \cite{alvarez_nmr_2010,alvarez_localization_2011},
we prepared a system of nuclear spins 1/2. Starting with uncorrelated
spins we let them evolve into clusters of correlated spins with increasing
size. By introducing a controlled perturbation to the Hamiltonian
that generates these clusters, we find that the size of the system
tends towards a limiting value determined by a dynamic equilibrium
\cite{alvarez_nmr_2010,alvarez_localization_2011}: if the cluster
size is initially larger than this equilibrium value, it decreases
under the effect of the perturbed Hamiltonian, and it increases while
its size is below the stationary value. The equilibrium size decreases
with increasing strength of the perturbation. 

The paper is organized as follows. Section II describes the quantum
simulator, the system and the initial state preparation. Section III
shows the quantum simulations. It is divided in two parts: III.A.
contains the unperturbed evolutions that drives the growth of the
clusters, and it desribes the technique for measuring the size of
the clusters. In section III.B., we discuss the perturbed evolutions,
we describe the perturbations and how we create them. In section IV,
we discuss the dynamical equilibrium with stationary cluster-size,
which is independent of the initial states with different cluster-sizes.
Lastly, section V gives the conclusions.

\section{\textit{\emph{The quantum simulator }}}

\subsection{System}

We consider a system of equivalent spins $I=1/2$ in the presence
of a strong magnetic field and subject to mutual dipole-dipole interaction.
The Hamiltonian of the system is 
\begin{align}
\widehat{\mathcal{H}} & =\widehat{\mathcal{H}}_{z}+\widehat{\mathcal{H}}_{dip},
\end{align}
where $\widehat{\mathcal{H}}_{z}=\omega_{z}\sum_{i}\hat{I}_{z}^{i}$
represents the Zeeman interaction with $\omega_{z}=\hbar\gamma B_{0}$
as the Larmor frequency, and
\begin{align}
\widehat{\mathcal{H}}_{dip} & =\frac{1}{2}\sum_{i<j}\frac{\vec{\mu}_{i}\cdot\vec{\mu}_{j}}{r_{ij}^{3}}-\frac{3\left(\vec{\mu}_{i}\cdot\vec{r}_{ij}\right)\left(\vec{\mu}_{j}\cdot\vec{r}_{ij}\right)}{r_{ij}^{5}}
\end{align}
is the dipolar interaction \cite{Slichter}, typically found also
in dipolar quantum gases \cite{Lahaye2009} and Rydberg atoms \cite{Saffman2010}
of growing interest in the context of quantum information science.
The dipoles are $\vec{\mu}_{i}=\hbar\gamma(\hat{I}_{x}^{i},\hat{I}_{y}^{i},\hat{I}_{z}^{i})$
with $\hat{I}_{x}^{i},\hat{I}_{y}^{i}\mbox{ and }\hat{I}_{z}^{i}$
the spin operators and $\vec{r}_{ij}$ is the distance vector between
$\vec{\mu}_{i}$ and $\vec{\mu}_{j}$. In the presence of a strong
magnetic field, ($\omega_{z}\gg d_{ij}$), it is possible to truncate
$\widehat{\mathcal{H}}_{dip}$ with respect to $\widehat{\mathcal{H}}_{z}$.
The part that does not commute has negligible effect on the evolution
of the system \cite{Slichter}, while the secular part can be written
as
\begin{align}
\widehat{\mathcal{H}}_{dd} & =\sum_{i<j}d_{ij}\left[2\hat{I}_{z}^{i}\hat{I}_{z}^{j}-(\hat{I}_{x}^{i}\hat{I}_{x}^{j}+\hat{I}_{y}^{i}\hat{I}_{y}^{j})\right].
\end{align}
The coupling constants are 
\begin{equation}
d_{ij}=\frac{1}{2}\frac{\gamma^{2}\hslash^{2}}{r_{ij}^{3}}\left(1-3\cos^{2}\theta_{ij}\right),\label{eq:dip_coupling}
\end{equation}
with $\theta{}_{ij}$ the angle between the vector $\vec{r}_{ij}$
and the magnetic field direction \cite{Slichter}. In a frame of reference
rotating at the Larmor frequency $\omega_{z}$ \cite{Slichter}, the
Hamiltonian of the spin system reduces to $\widehat{\mathcal{H}}_{dd}$.
This kind of Hamiltonians can be also simulated with quantum gases
\cite{Lahaye2009}. 

\begin{figure}
\includegraphics[width=1\columnwidth,height=0.8\columnwidth]{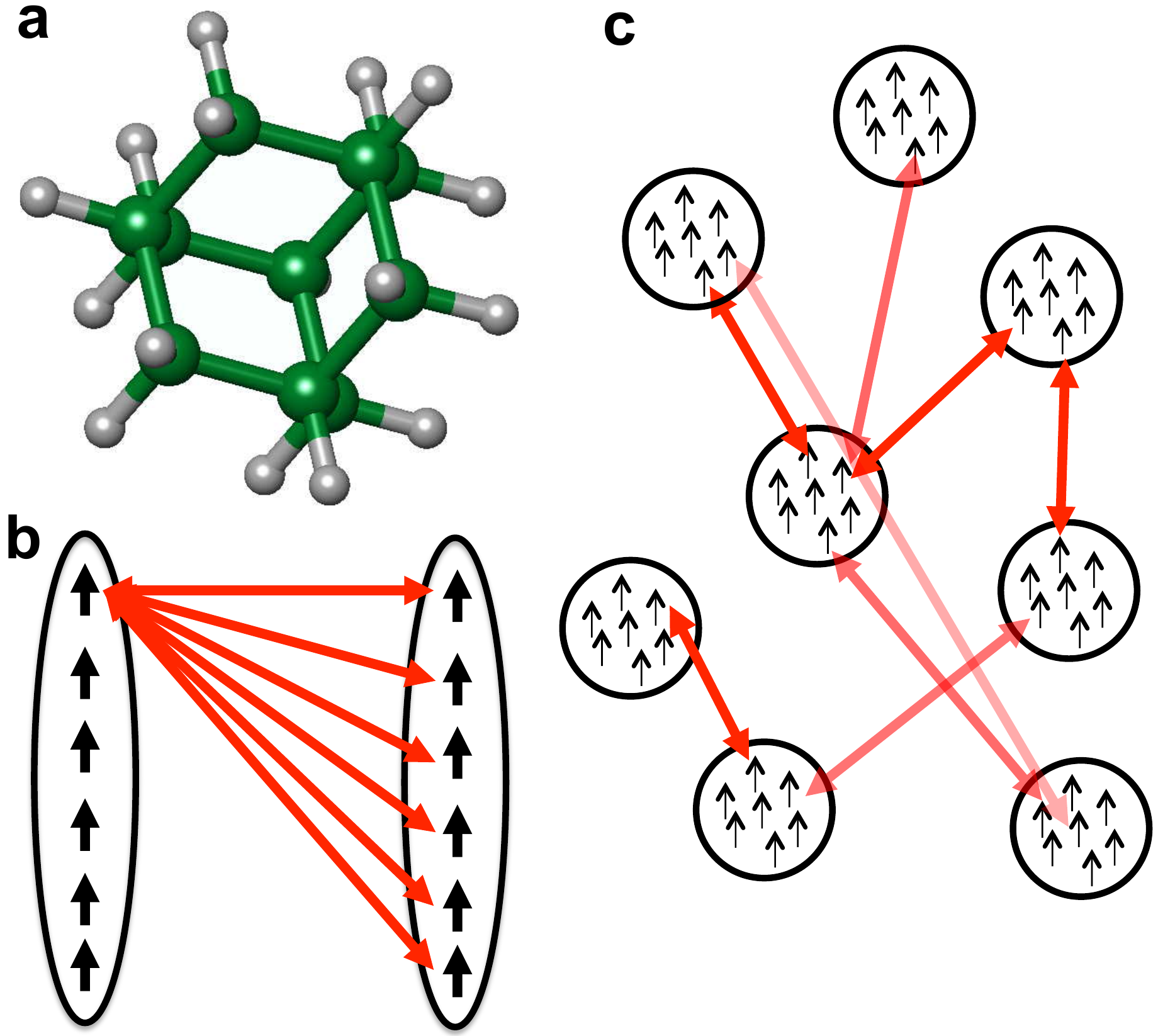}

\caption{\label{fig:systemscheme}(Color online) Spin system. (a) Adamantane
molecule with 16 protons (small gray spheres). The big green spheres
are mostly $^{12}$C and $^{13}$C in natural abundance (1.1\%$%%
\ensuremath{}$). (b) The intramolecular interactions are averaged to zero due to
very fast molecular tumbling, but the intermolecular interactions
average to a non-zero value that depends of the distance between the
molecules. The spins interacts with spins of the other molecule with
the same averaged coupling strength, as shown with arrows. (c) Schematic
representation of the interactions between the molecules. The color
tones of the arrows represent the variation of the coupling strength
with the intermolecular distance, which varies as $1/r_{ij}^{3}$.}
\end{figure}

In our system, the spins are the protons of polycrystalline adamantane
and we performed all experiments on a home-built solid state NMR spectrometer
with a $^{\text{1}}$H resonance frequency of $\omega_{z}=300$ MHz
in Dortmund. As shown in Fig.\ \ref{fig:systemscheme}, the adamantane
molecule is nearly spherical and contains 16 protons. The molecules
tumble rapidly and isotropically in the solid phase. This fast tumbling
averages the intramolecular couplings to zero, but the interaction
between the molecules remains. However, the couplings between molecules
are averaged to a nonzero value that depends only on the relative
position of the molecules to which the spins belong. They are not
isotropic, and they have the normal orientation dependence of dipolar
couplings of Eq. (\ref{eq:dip_coupling}), but the distance is between
the positions of the molecules, not of the nuclei. ``Position'' would
in fact not be the center of mass of the molecules, but an effective
position that is the result of the averaging process. Figure \ref{fig:systemscheme}b
shows a scheme of the interaction between two molecules, where the
spins do not interact with spins within the molecules but they interact
with all spins of the neighbor molecules. All coupling strength are
averaged to the same value. However, in Fig. \ref{fig:systemscheme}c,
the coupling strength between molecules depends of their separation,
as shown by arrows with different color tones, and on their the polar
angle $\theta{}_{ij}$ with respect to the magnetic field. The randomness
of the dispersion of the distance vector $\vec{r}$ between molecules
will be the source of disorder on $d_{ij}$. The molecules are in
a face-centered-cubic lattice, where each adamantane molecule has
$12$ first neighbor interactions with a distance of $6.6\textrm{\AA}$,
then $6$ second neighbors with a distance of $9.34\textrm{\AA}$,
then $16$ at a distance of $11.4\textrm{\AA}$ and etc. The resonance
width of the NMR resonance line, is $7.9$ kHz.

\subsection{Initial state preparation}

We perform quantum simulations starting from the high-temperature
thermal equilibrium \cite{Slichter}. Using the notation $\hat{I}_{z}=\sum_{i}\hat{I}_{z}^{i}$,
we can write the thermal equilibrium state as 
\begin{eqnarray}
\rho_{0} & = & \exp\left\{ -\frac{\hbar\omega_{z}}{k_{\mathrm{B}}T}\hat{I}_{z}\right\} /\mathrm{Tr}\left\{ \exp\left\{ -\frac{\hbar\omega_{z}}{k_{\mathrm{B}}T}\hat{I}_{z}\right\} \right\} \\
 & \approx & \left(\hat{1}+\frac{\hbar\omega_{z}}{k_{\mathrm{B}}T}\hat{I}_{z}\right)/\mathrm{Tr}\left\{ \hat{1}\right\} .
\end{eqnarray}
 It is convenient to exclude the unity operator $\hat{1}$ since it
does not evolve in time and does not contribute to the observable
signal. The resulting state is $\hat{\rho}_{0}\propto\hat{I}_{z}$.
In this state, the spins are uncorrelated and the density operator
commutes with the Hamiltonian $\widehat{\mathcal{H}}_{dd}$. In order
to prepare a different initial state, we wait a time longer that $T_{1}$
to reinitialize the system state to $\rho_{0}$.

\section{Quantum simulations}

\subsection{Unperturbed evolution}

\subsubsection{Generating clusters}

\begin{figure}
\includegraphics[bb=0bp 0bp 316bp 171bp,width=0.7\columnwidth]{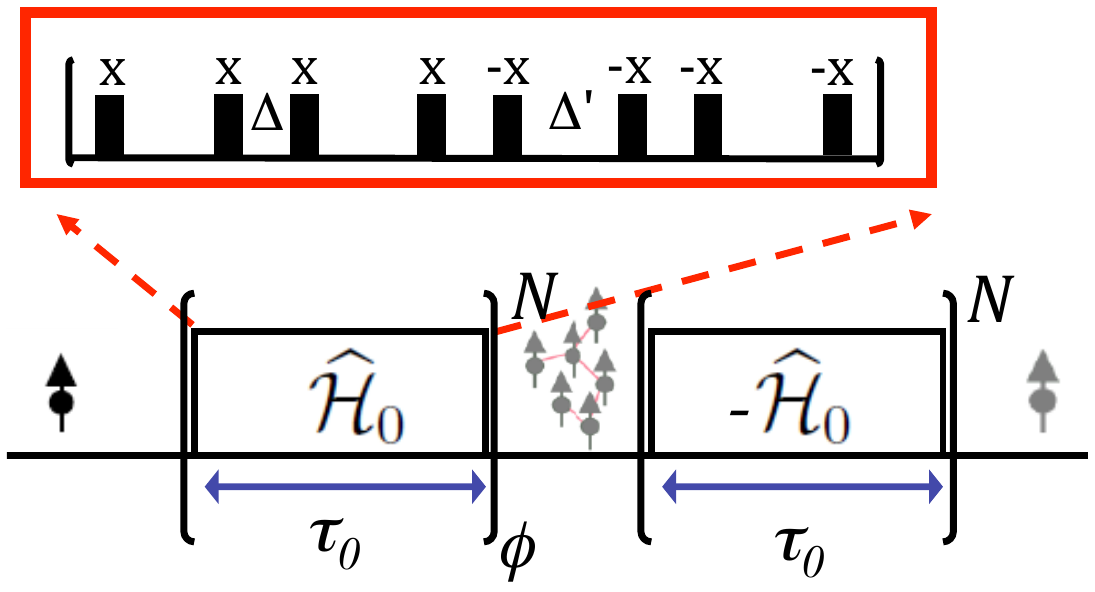}

\caption{(Color online) Pulse sequence for the quantum simulations. (a) The
effective Hamiltonian $\widehat{\mathcal{H}}_{0}$ is generated by
the periodic sequence of $\text{\ensuremath{\pi}/2}$ pulses. The
upper part of the figure shows the basic cycle, where $\Delta^{\prime}=2\Delta+\tau_{p}$,
$\Delta=2\mu$s and $\tau_{p}=2.8\mu$s is the $\text{\ensuremath{\pi}/2}$
pulse duration \cite{Baum1985}. The cycle time is then $\tau_{0}=57.6\mu$s. }

\label{Flo:NMRseqH0-1} 
\end{figure}

The initial state $\rho_{0}$ of the uncorrelated spins commutes with
$\widehat{\mathcal{H}}_{dd}$. Therefore, to generate spin clusters
we use an NMR method developed by Pines and coworkers \cite{5105,Baum1985}.
It is based on generating an average Hamiltonian $\widehat{\mathcal{H}}_{0}$
that does not commute with the thermal equilibrium state

\begin{eqnarray}
\widehat{\mathcal{H}}_{0} & = & -\sum_{i<j}d_{ij}\left[\hat{I}_{x}^{i}\hat{I}_{x}^{j}-\hat{I}_{y}^{i}\hat{I}_{y}^{j}\right].\label{flip-flip}\\
 & = & -\frac{1}{4}\sum_{i<j}d_{ij}\left[\hat{I}_{+}^{i}\hat{I}_{+}^{j}-\hat{I}_{-}^{i}\hat{I}_{-}^{j}\right].
\end{eqnarray}
This Hamiltonian drives an evolution that converts the thermal initial
state into clusters of correlated spins whose density operator contains
terms of the form $\hat{I}_{u}^{i}...\hat{I}_{v}^{j}\hat{I}_{w}^{k}\left(u,v,w=x,y,z\right)$,
where the indexes $i,j,k$ identify the spins involved in the given
cluster. The cluster-size $K$ corresponds to the number of terms
in this product, which is equal to the number of spins. The cluster
size is related to the volume occupied by those spins. Experimentally,
the Hamiltonian $\widehat{\mathcal{H}}_{0}$ is generated with the
pulse sequence \cite{5105,Baum1985} shown in the upper part of Fig.
\ref{Flo:NMRseqH0-1}.

\begin{figure}
\includegraphics[width=1\columnwidth]{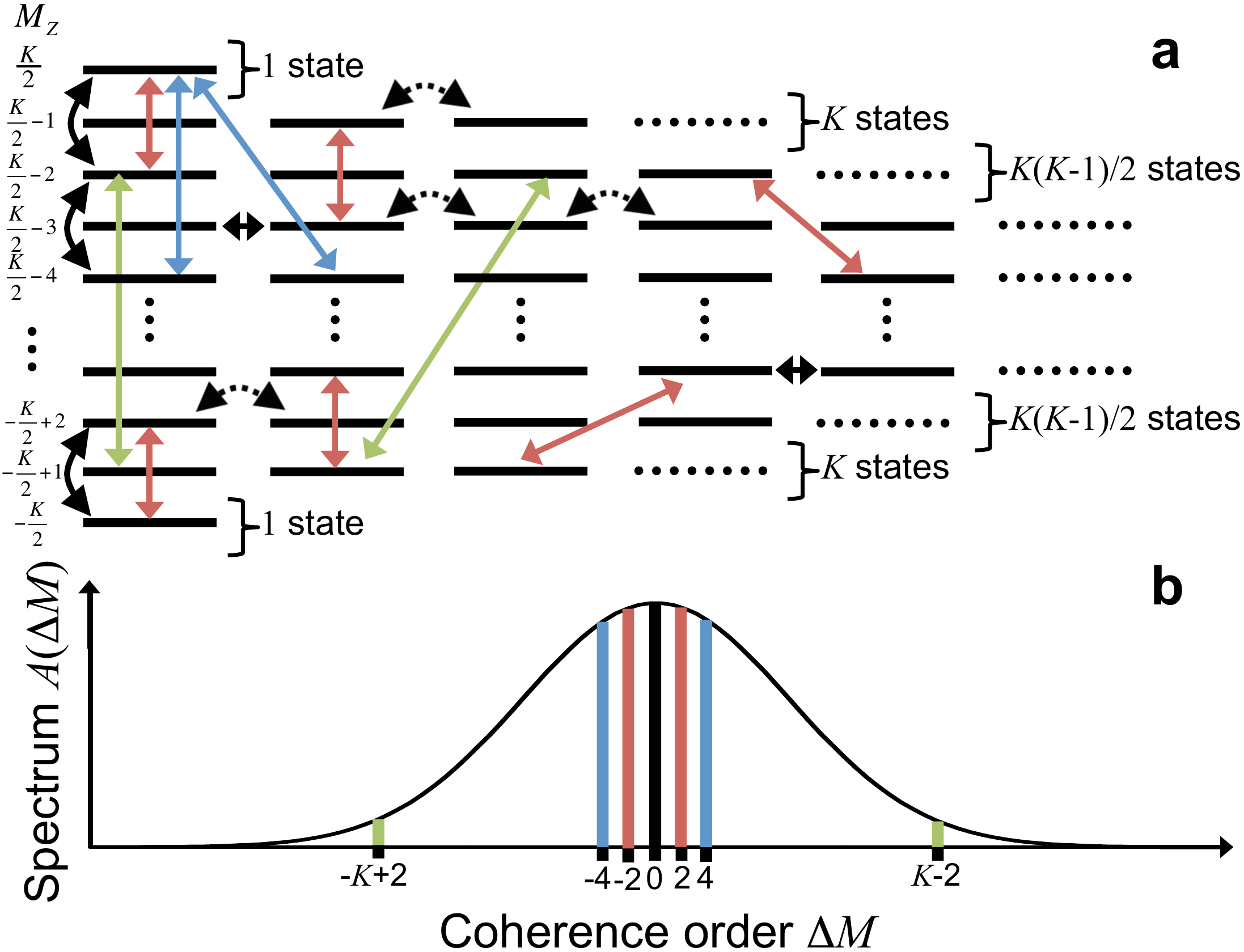}

\caption{\label{fig:levelschemesAm}(Color online) Energy level scheme for
a cluster of $K$ spins. (a) The different rows correspond to Zemman
states $\left|\alpha_{1},\alpha_{2},...,\alpha_{K}\right\rangle $
($\alpha_{i}=\uparrow,\downarrow$) with different energy determined
by the quantum number $M_{z}$. The degeneracy of the levels in a
row is given on the rhs of each row. The solid curved arrows show
those transitions induced by the Hamiltonian $\mathcal{H}_{0}$ that
do not conserve $M_{z}$. The dotted curved arrows show the effect
of the $M_{z}$-conserving dipolar Hamiltonian $\mathcal{H}_{dd}$.
The straight colored arrows show the possible coherences generated
by $\mathcal{H}_{0}$. (b) The number of coherences of a cluster of
size $K$ are plotted as a function of $\Delta M$. The colored bars
gives those numbers and their color code corresponds to thatof panel
(a).}
\end{figure}

In the usual computational or Zeeman basis $\left|\alpha_{1},\alpha_{2},...,\alpha_{K}\right\rangle $
($\alpha_{i}=\uparrow,\downarrow$) for a system of $K$ spins, we
write the states as $\left|M_{z},n_{M}\right\rangle $ where $M_{z}$
is the total magnetic quantum number, \emph{i.e.}, $\hat{I}_{z}\left|M_{z},n_{M}\right\rangle =M_{z}\left|M_{z},n_{M}\right\rangle $,
and $n_{M}$ distinguishes different states with the same $M_{z}$.
Figure \ref{fig:levelschemesAm}a shows a summary of these states,
whose degeneracy is $\max\left\{ n_{M}\right\} =K!/\left[\left(K/2-M_{z}\right)!\left(K/2+M_{z}\right)!\right]$.

\begin{figure}
\includegraphics[width=1\columnwidth,height=0.6\columnwidth]{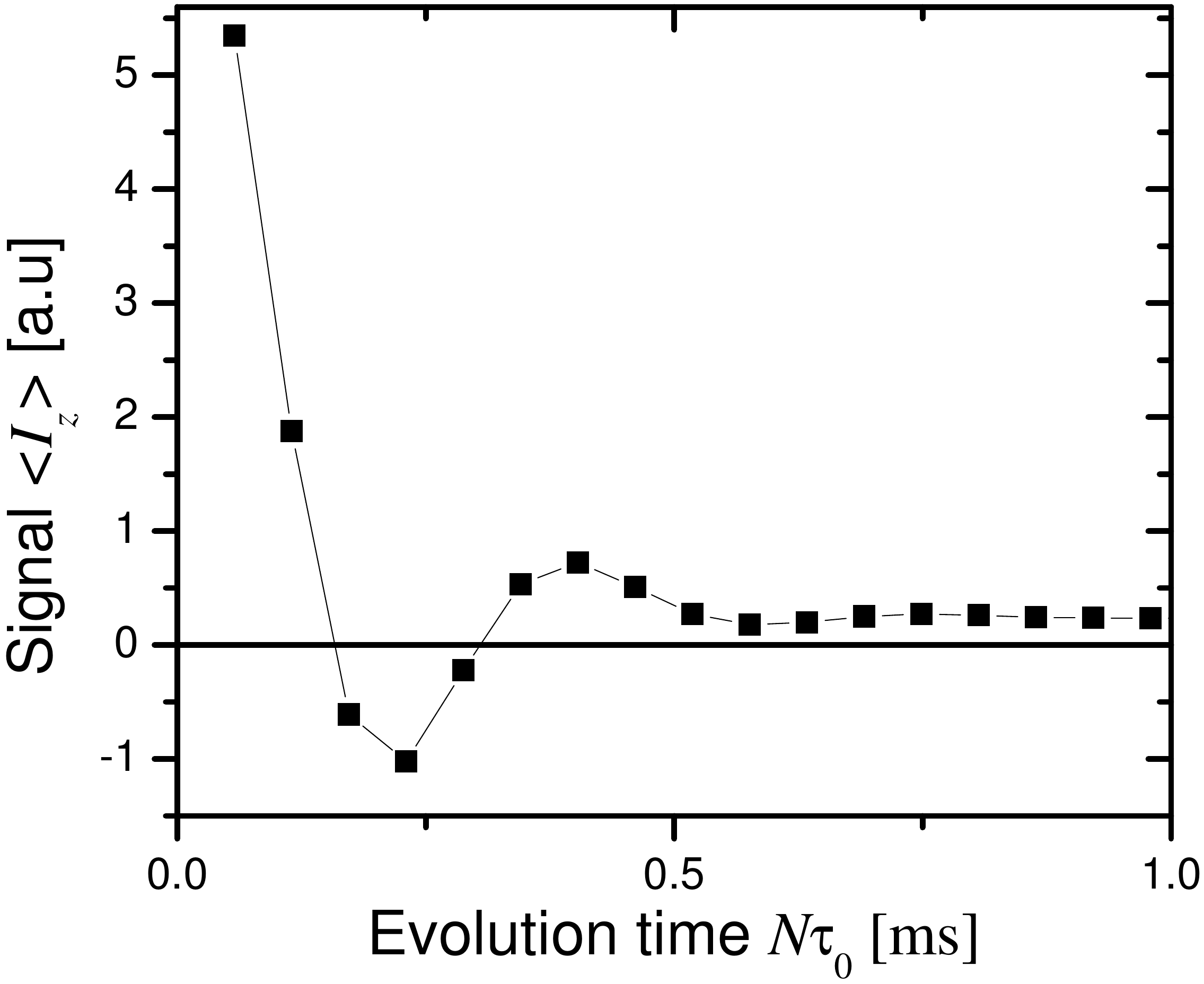}

\caption{\label{fig:Fwddynamics}Time evolution driven by $\mathcal{H}_{0}$
of the total magnetization $\left\langle I_{z}(t)\right\rangle $
of the system as a function of time. The black squares are the unperturbed
evolution, while the remaining symbols are perturbed evolutions according
to the legend.}
\end{figure}

The Hamiltonian $\widehat{\mathcal{H}}_{0}$ flips simultaneously
two spins, which are separated in space and have the same orientation.
Accordingly, the $z$-component of the magnetization $M_{z}$ changes
by $\Delta M_{z}=\pm2.$ This is shown with a curved solid arrow in
Fig. \ref{fig:levelschemesAm}a. At the same time, the number $K$
of correlated spins changes by $\Delta K=\pm1$. Therefore, starting
from the thermal equilibrium state, the evolution generates a density
operator where only elements $\rho_{ij}$ with $\Delta M=M_{z}(i)-M_{z}(j)=2n,\, n=0,1,2\dots$
are populated. Such elements $\rho_{ij}$ are called $\Delta M$ quantum
coherences and they are represented by colored straight arrows in
Fig. \ref{fig:levelschemesAm}a. The different colors represent different
multiple-quantum coherence (MQC) orders $\Delta M$. Off-diagonal
elements of the density matrix with $\Delta M=0$ represent zero-quantum
coherences and diagonal elements correspond to populations. Then,
a MQC spectrum $A(\Delta M)$ can be described by the number of coherences
of the density matrix for a given $\Delta M$. A typical MQC spectrum
is shown in Fig. \ref{fig:levelschemesAm}b. The initial state $\rho_{0}$
is diagonal and then $A(\Delta M)\neq0$ only for $\Delta M=0$. However,
as time evolves, different spins interact with each other and other
coherence orders are excited. Then $A(\Delta M)$ starts to spread
as a manifestation of the increasing cluster-size. If we measure the
evolution of the operator $I_{z}$ as a function of time, $\left\langle I_{z}(t)\right\rangle =\mathrm{Tr}\left\{ I_{z}\rho(t)\right\} $,
its expectation value decays as a consequence of the excitation of
the coherences of the density matrix that do not contribute to the
observable $\left\langle I_{z}(t)\right\rangle $. The black squares
in Fig. \ref{fig:Fwddynamics} show the evolution of $\left\langle I_{z}(t)\right\rangle $
driven by the Hamiltonian $\widehat{\mathcal{H}}_{0}$. 
We can see a fast decay on a time scale of $\approx100\,\mu$s, which
is followed by a quantum beat at about 400$\mu$s. Then the signal
saturates.

\subsubsection{Measuring cluster sizes}

\label{sub:MeasureClusterSize}

To determine the average number of correlated spins in the generated
clusters, we use the NMR technique developed by Baum \emph{et al.}
\cite{Baum1985}. In a system of $K$ spins, the number of transitions
with a given $\Delta M$ follows a binomial distribution 
\begin{equation}
n\left(\Delta M,K\right)=\frac{\left(2K\right)!}{\left(K+\Delta M\right)!\left(K\text{\textminus}\Delta M\right)!}.
\end{equation}
For $K\gg1,$ the binomial distribution can be approximated with a
Gaussian
\begin{equation}
n\left(\Delta M,K\right)\propto\exp\left(\text{\textminus}\frac{\Delta M^{2}}{K}\right),
\end{equation}
whose half width at $e^{-1}$ is $\sigma=\sqrt{K}$ . Thus, the width
of the MQC spectrum reflects the cluster-size. We can determine the
effective size of the spin clusters in a given state by measuring
the distribution of the MQCs of its density operator $\rho$ as a
function of the coherence order $\Delta M$. They can be distinguished
experimentally by rotating the system around the $z-$axis: a rotation
$\hat{\phi}_{z}=e^{-i\phi\hat{I}_{z}}$ by $\phi$ changes the density
operator to

\begin{equation}
\hat{\rho}\left(\phi\right)=\hat{\phi}_{z}\hat{\rho}\hat{\phi}_{z}^{\dagger}=\sum_{\Delta M}\hat{\rho}_{\Delta M}^{{}}e^{i\Delta M\phi},\label{eq:rhophi}
\end{equation}
where $\hat{\rho}_{\Delta M}$ contains all the elements of the density
operator involving coherences of order $\Delta M$. 

\begin{figure}
\begin{centering}
\includegraphics[width=1\columnwidth]{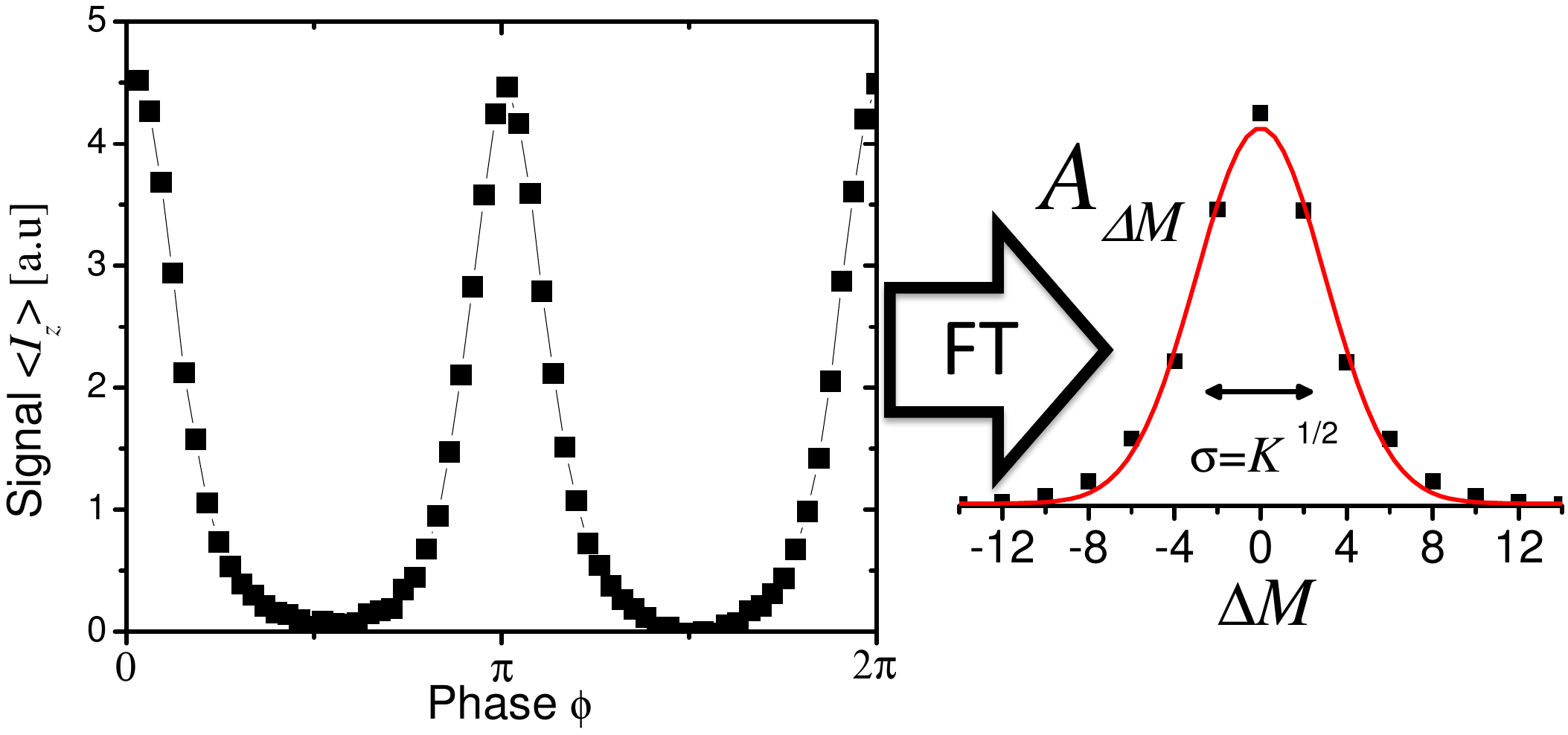}
\par\end{centering}

\caption{\label{fig:signalvsphi}(Color online) Scheme for determining the
MQC spectrum based on the sequence of Fig. \ref{Flo:NMRseqH0-1}.
The left panel is the measured $\left\langle I_{z}\right\rangle $
signal as a function of the encoding phase $\phi$ for $N\tau_{0}=230.4\mu$s.
After doing a Fourier transform the MQC spectrum $A(\Delta M)$ is
obtained (right panel). In the left panel, the black squares are the
unperturbed signal ($p=0)$, while the other colored symbols are the
signal observed for a given perturbation strength $p$.}
\end{figure}

By following the sequence of Fig. \ref{Flo:NMRseqH0-1}, the system
evolution is first described by an evolution period of duration $N\tau_{0}$
under the Hamiltonian $\left(\mathcal{\widehat{H}}_{0}\right)_{\phi}=\hat{\phi}_{z}\mathcal{\widehat{H}}_{0}\hat{\phi}_{z}^{\dagger}$,
\emph{i.e.}, 
\begin{align}
\hat{\rho}_{0}\xrightarrow{\left(\mathcal{\widehat{H}}_{0}\right)_{\phi}N\tau_{0}}\hat{\rho}_{\phi}\left(N\tau_{0}\right) & =\hat{\phi}_{z}\hat{\rho}\left(N\tau_{0}\right)\hat{\phi}_{z}^{\dagger}\nonumber \\
 & =\hat{\phi}_{z}e^{-i\widehat{\mathcal{H}}_{0}N\tau_{0}}\hat{\rho}_{0}e^{i\mathcal{\widehat{H}}_{0}N\tau_{0}}\hat{\phi}_{z}^{\dagger}\nonumber \\
 & =\sum_{\Delta M}\hat{\phi}_{z}\hat{\rho}_{\Delta M}^{{}}\left(N\tau_{0}\right)\hat{\phi}_{z}^{\dagger}\nonumber \\
 & =\sum_{\Delta M}\hat{\rho}_{\Delta M}^{{}}\left(N\tau_{0}\right)e^{i\Delta M\phi}.
\end{align}
The next part of the sequence of Fig. \ref{Flo:NMRseqH0-1} is an
evolution of the same duration $N\tau_{0}$ under $-\mathcal{\widehat{H}}_{0}$.
This causes an evolution backward in time that gives the following
density operator at the end of the sequence
\begin{multline}
\xrightarrow{\left(-\mathcal{\widehat{H}}_{0}\right)N\tau_{0}}\hat{\rho}_{f}\left(2N\tau_{0}\right)=e^{i\widehat{\mathcal{H}}_{0}N\tau_{0}}\hat{\rho}_{\phi}\left(N\tau_{0}\right)e^{-i\widehat{\mathcal{H}}_{0}N\tau_{0}}\\
=\sum_{\Delta M}\left[e^{i\mathcal{\widehat{H}}_{0}N\tau_{0}}\hat{\rho}_{\Delta M}^{{}}\left(N\tau_{0}\right)e^{-i\mathcal{\widehat{H}}_{0}N\tau_{0}}\right]e^{i\Delta M\phi}.
\end{multline}
If $\hat{I}_{z}$ is the NMR observable, then the signal becomes 
\begin{align}
\left\langle \hat{I}_{z}\right\rangle \left(\phi,N\tau_{0}\right) & =\mbox{Tr}\left\{ \hat{I}_{z}\hat{\rho}_{f}\left(2N\tau_{0}\right)\right\} \nonumber \\
 & =\mathrm{Tr}\left\{ e^{-i\mathcal{\widehat{H}}_{0}N\tau_{0}}\hat{\rho}_{0}e^{i\mathcal{\widehat{H}}_{0}N\tau_{0}}\hat{\rho}_{\phi}\left(N\tau_{0}\right)\right\} \nonumber \\
 & =\mathrm{Tr}\left\{ \hat{\rho}\left(N\tau_{0}\right)\hat{\rho}_{\phi}\left(N\tau_{0}\right)\right\} \nonumber \\
 & =\sum_{\Delta M}\mbox{\ensuremath{e^{i\phi\Delta M}}Tr}\left\{ \hat{\rho}_{\Delta M}^{2}\left(N\tau_{0}\right)\right\} \label{eq:unperturbed_signal-1}\\
 & =\sum_{\Delta M}\mbox{\ensuremath{e^{i\phi\Delta M}}}A\left(\Delta M\right),
\end{align}
where $A(\Delta M)$ are the amplitudes of the MQ spectrum shown in
Fig. \ref{fig:levelschemesAm}b. To extract these amplitudes from
the experimental data, we measure the signal $\left\langle \hat{I}_{z}\right\rangle \left(\phi,N\tau_{0}\right)$
as a function of $\phi$ at a fixed time $N\tau_{0}$ and then perform
a Fourier transform with respect to $\phi$ as shown schematically
in Fig. \ref{fig:signalvsphi} (Black squares). The cluster size is
then determined by the half-width at $e^{-1}$, $\sigma=\sqrt{K}$,
of $A(\Delta M)$. 

\subsubsection{Growth of the clusters}

\begin{figure}
\begin{centering}
\includegraphics[width=1\columnwidth]{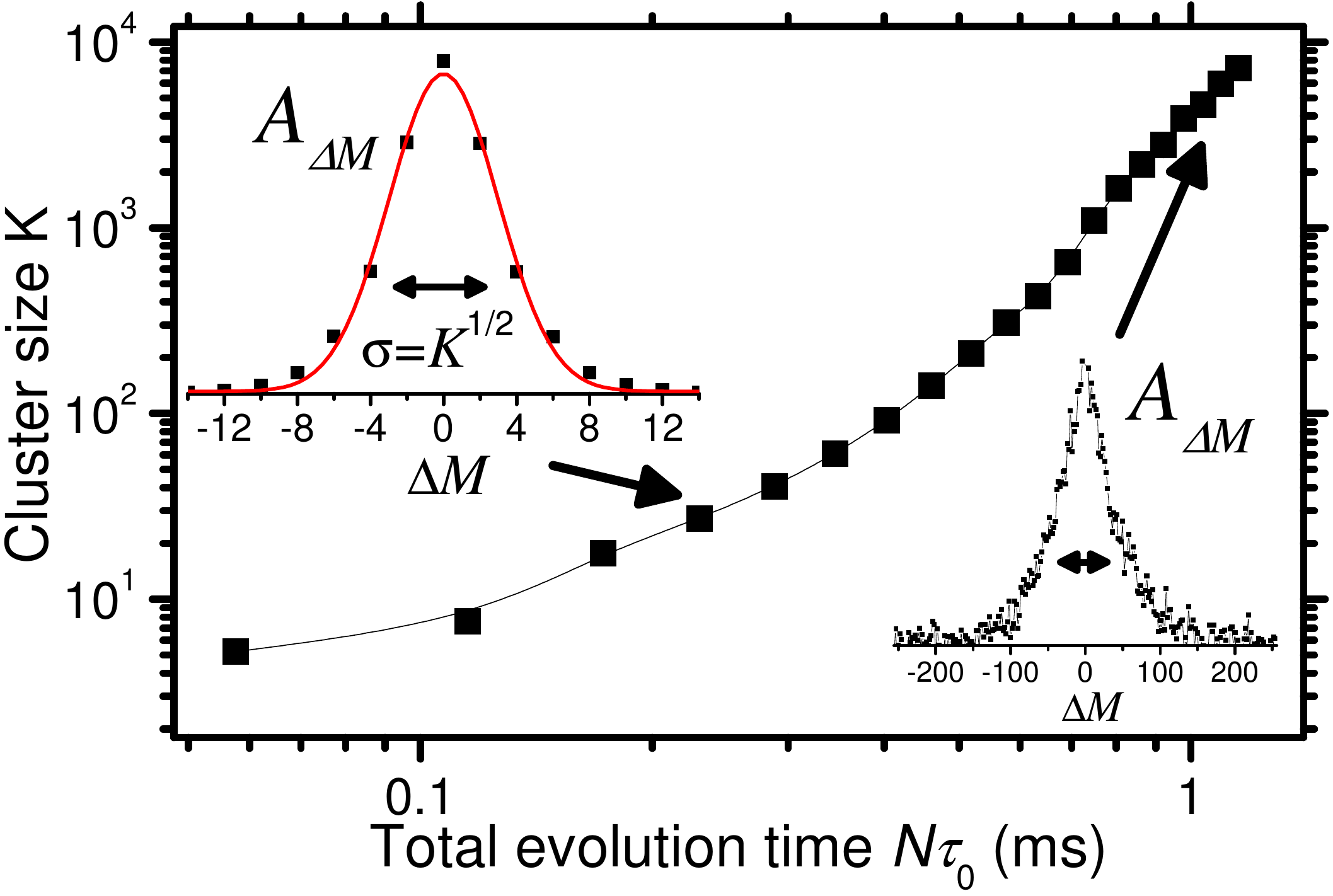}
\par\end{centering}

\caption{(Color online) Time evolution of the cluster size of correlated spins
with the unperturbed Hamiltonian $\widehat{\mathcal{H}}_{0}$ (black
squares). Distributions of the squared amplitudes $A_{\Delta M}$
of density operator components as a function of the coherence order
$\Delta M$ are shown for two different cluster sizes. }

\label{fig:Growth} 
\end{figure}

Figure \ref{fig:Growth} shows the time evolution of the measured
cluster size $K\left(N\tau_{0}\right)$ as a function of the total
evolution time $N\tau_{0}$. For the unperturbed Hamiltonian, the
cluster size appears to grow indefinitely \cite{alvarez_nmr_2010,alvarez_localization_2011}.
The figure also shows two examples of the $A(\Delta M)$ distributions
at different times. We can see that for long times $K\propto\left(N\tau_{0}\right)^{4.3}$.
Assuming that the cluster-size $K$ is associated with the number
of spins inside a volume, it is seen that the associated length grow
faster than normal diffusion, where its growing curve would be expected
to be $\propto\left(N\tau_{0}\right)^{3/2}$.

This evolution can be reversed completely by changing the Hamiltonian
from $\widehat{\mathcal{H}}_{0}$ to $-\widehat{\mathcal{H}}_{0}$.
Experimentally, this is achieved by shifting the phase of all RF pulses
by $\pm\pi/2$ \cite{5105}. The signal $\left\langle I_{z}\right\rangle \left(\phi,N\tau_{0}\right)$
at the end of the sequence of Fig. \ref{Flo:NMRseqH0-1} is a time
reversal echo for $\phi=0$. This means that under ideal conditions
$\sum_{M}A(\Delta M,N\tau_{0})=\mathrm{const}$ and we will write
$E\left(N\tau_{0}\right)$ for this quantity. The indefinite growth
of the cluster size, as well as the complete reversibility of the
time evolution are no longer possible if the effective Hamiltonian
deviates from the ideal form (\ref{flip-flip}).

\subsection{Effect of perturbations}

\subsubsection{Intrinsic perturbations}

Experimentally, the Hamiltonian $\widehat{\mathcal{H}}_{0}$ is generated
as an effective Hamiltonian by the pulse sequence of Fig. \ref{Flo:NMRseqH0-1}.
Because of experimental imperfections, it always deviates from the
ideal Hamiltonian (\ref{flip-flip}). As a result, the actual dynamics
deviates from the ideal one and, in particular, we cannot invert exactly
the perturbed Hamiltonian and thus reverse the time evolution perfectly.
The quantity $\sum_{\Delta M}A(\Delta M)$ is no longer conserved,
but decays with increasing evolution time. 

The ideal form of the effective Hamiltonian (\ref{flip-flip}) can
only be created if the dipolar couplings $d_{i,j}$ are time independent
and the pulses are ideal and rotate globally all the spins. However,
if these couplings are time dependent, or the pulses are not ideal,
the effective Hamiltonian (averaged over the pulse cycle) contains
additional terms. We have partly characterized the spectral density
of local spin-spin fluctuations driven by $\widehat{\mathcal{H}}_{dd}$
in Ref. \cite{Alvarez2011,Alvarez2010c} and its correlation time
was about $\tau_{d}=110\mu$s. Since the imperfection on $\widehat{\mathcal{H}}_{0}$
are correlated with the fluctuations driven by $\widehat{\mathcal{H}}_{dd}$,
we can use $\tau_{d}$ to estimate the correlation time of the fluctuations
of the $\widehat{\mathcal{H}}_{0}$ imperfections.

In Fig. \ref{fig:Growth} the cluster size grows faster for times
lower than $\tau_{d}$. After $\tau_{d}$ the cluster size growth
reduces speed and seems to start growing exponentially \cite{alvarez_nmr_2010,alvarez_localization_2011}.
After 1ms, the growing law again change its behavior and the cluster
size grows as a power law \cite{alvarez_nmr_2010,alvarez_localization_2011}.
We cannot contrast this regime with the spectral density determined
in Ref. \cite{Alvarez2011} because it was not determined for the
corresponding range of frequency fluctuations. While all these regimes
cannot be rigorously determined, there is a fact that the cluster-size
keeps growing for long time and faster than normal diffusion. This
`super-diffusion' may be a result of the long range nature of the
dipolar interaction \cite{Metzler2000,Mercadier2009}.

\begin{figure}
\includegraphics[width=1\columnwidth,height=0.5\columnwidth]{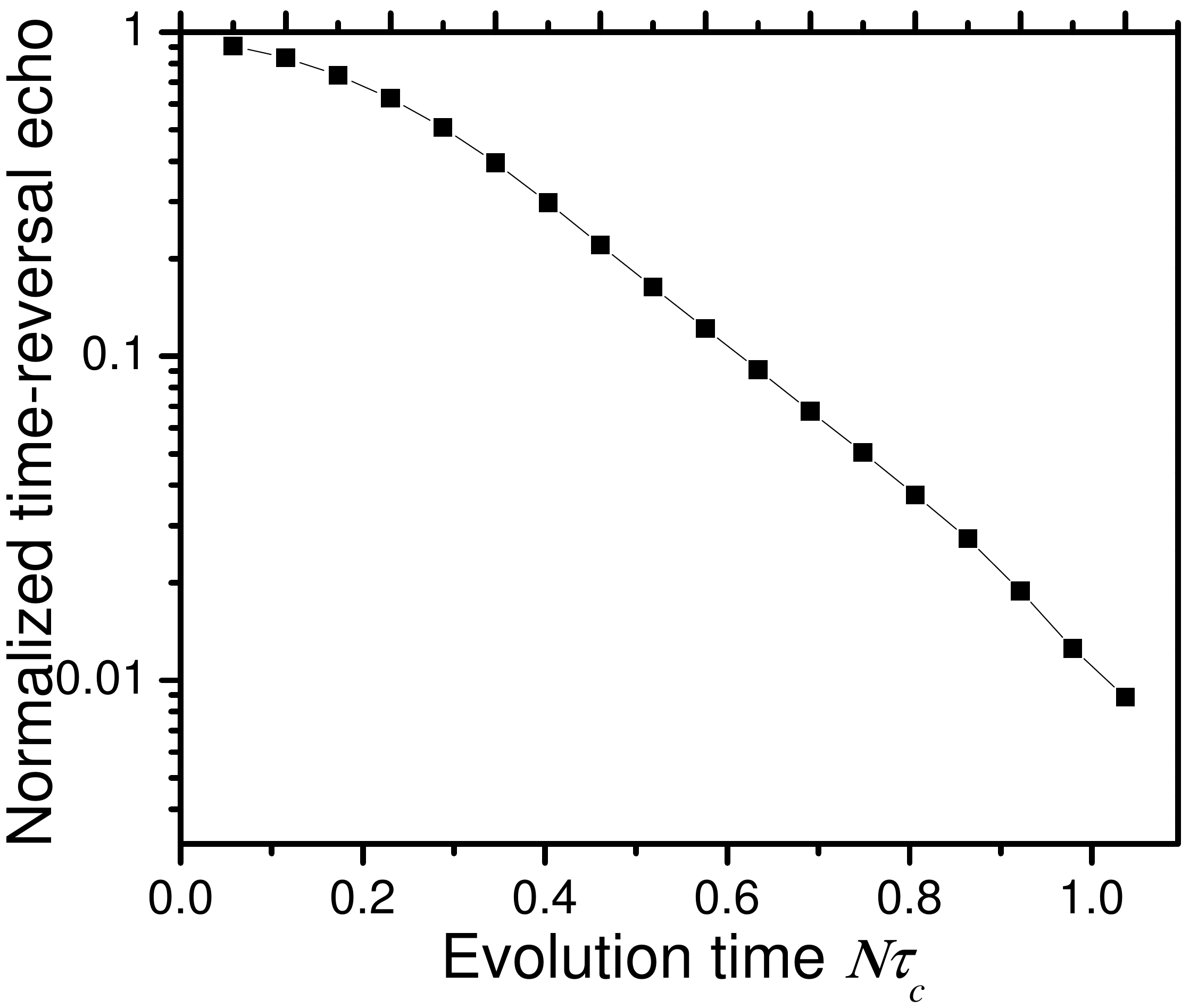}

\caption{\label{fig:echodecay}Time-reversal echo probability. The black squares
are the unperturbed ($p=0$) echo decay $E(N_{0}\tau_{0})$, measured
with the sequence of Fig. \ref{Flo:NMRseqH0-1}, as a function of
the evolution time $N_{0}\tau_{0}$. The left axis gives the time-reversal
probability normalized with respect to the signal at $N=0$. }
\end{figure}

In previous works \cite{alvarez_nmr_2010,alvarez_localization_2011}
and in Fig. \ref{fig:Growth}, we determined the cluster growth by
isolating the intrinsic decay generated by an imperfect effective
Hamiltonian (\ref{flip-flip}). The imperfection effects are manifested
in the overall decrease of the echo signal $E\left(N\tau_{0}\right)$
by normalizing the MQ spectra such that the total signal $\sum_{\Delta M}A(\Delta M)$
for $\phi=0$ is constant in time. Now, we measured the decay of $E\left(N\tau_{0}\right)=\sum_{\Delta M}A(\Delta M,N\tau_{0})$.
The results are shown in Fig. \ref{fig:echodecay}. If we consider
the imperfections on $\widehat{\mathcal{H}}_{0}$, the effective Hamiltonian
during the first part of the sequence of Fig. \ref{Flo:NMRseqH0-1}a
will be $\widehat{\mathcal{H}}_{fwd}=\widehat{\mathcal{H}}_{0}+\widehat{\mathcal{H}}_{e}$,
and $\widehat{\mathcal{H}}_{bwd}=-\widehat{\mathcal{H}}_{0}+\widehat{\mathcal{H}}_{e}^{\prime}$
during the time reversed part, where $\widehat{\mathcal{H}}_{e}$
is an average error Hamiltonian. The echo decay is then $E(N_{0}\tau_{0})=\mathrm{Tr}\left\{ \hat{\rho}^{\mathcal{H}_{fwd}}\left(N\tau_{0}\right)\hat{\rho}^{\mathcal{H}_{bwd}}\left(N\tau_{0}\right)\right\} $
and it quantifies the time reversal probability as a kind of Loschmidt
echo \cite{PhysicaA,JalPas01,Rhim1970}.

The echo decay $E(N_{0}\tau_{0})$ shown in Fig. \ref{fig:echodecay}
starts as a Gaussian for times shorter than $N_{0}\tau_{0}\lesssim288\mu$s$\approx2.6\,\tau_{d}$.
For longer times it decays exponentially until $\approx920\mu$s where
a different decay law arises. These transitions between different
decay laws seems to be correlated with the growing law transitions
discussed in the previous parragraphs. This resembles the typical
behavior of nuclear spins of a solution diffusing in a inhomogeneous
magnetic field with a given standard deviation \cite{Klauder1962}
or in restricted spaces in the presence of a magnetic field gradient
\cite{Kennan1994}. If the spins only interact with the magnetic field,
due to the diffusion process, they feel a different magnetic field
at different times causing dephasing of the spin signal. The frequency
fluctuations have a correlation time given by the time needed to explore
the standard deviation of the changes of the magnetic field that is
related with the inhomogenity of the magnetic field or to the restriction
length. By applying a spin-echo sequence -a time reversion of the
spin precession by an inversion pulse that changes the sign of the
magnetic field interaction- one can partly reverse the effects of
the diffusion-driven dephasing \cite{Hahn1950,Carr1954}. The echo
sequence is analogous to the one of Fig. \ref{Flo:NMRseqH0-1}. If
the inversion pulse inducing the time reversal is applied at times
shorter than the correlation time of the frequency fluctuations of
the spins, the signal decay depends of the spatial displacement of
the spins and the signal decays faster as the time passes because
the dephasing rate increases with the displacement length. However,
for times longer than the correlation time of the frequency fluctuations
the decay rate becomes independent of the displacement and becomes
exponential, similar to the time reversal echo behavior shown in Fig.
\ref{fig:echodecay}.

\subsubsection{Controlled perturbation}

\begin{figure}
\includegraphics[bb=30bp 235bp 570bp 330bp,clip,width=1\columnwidth]{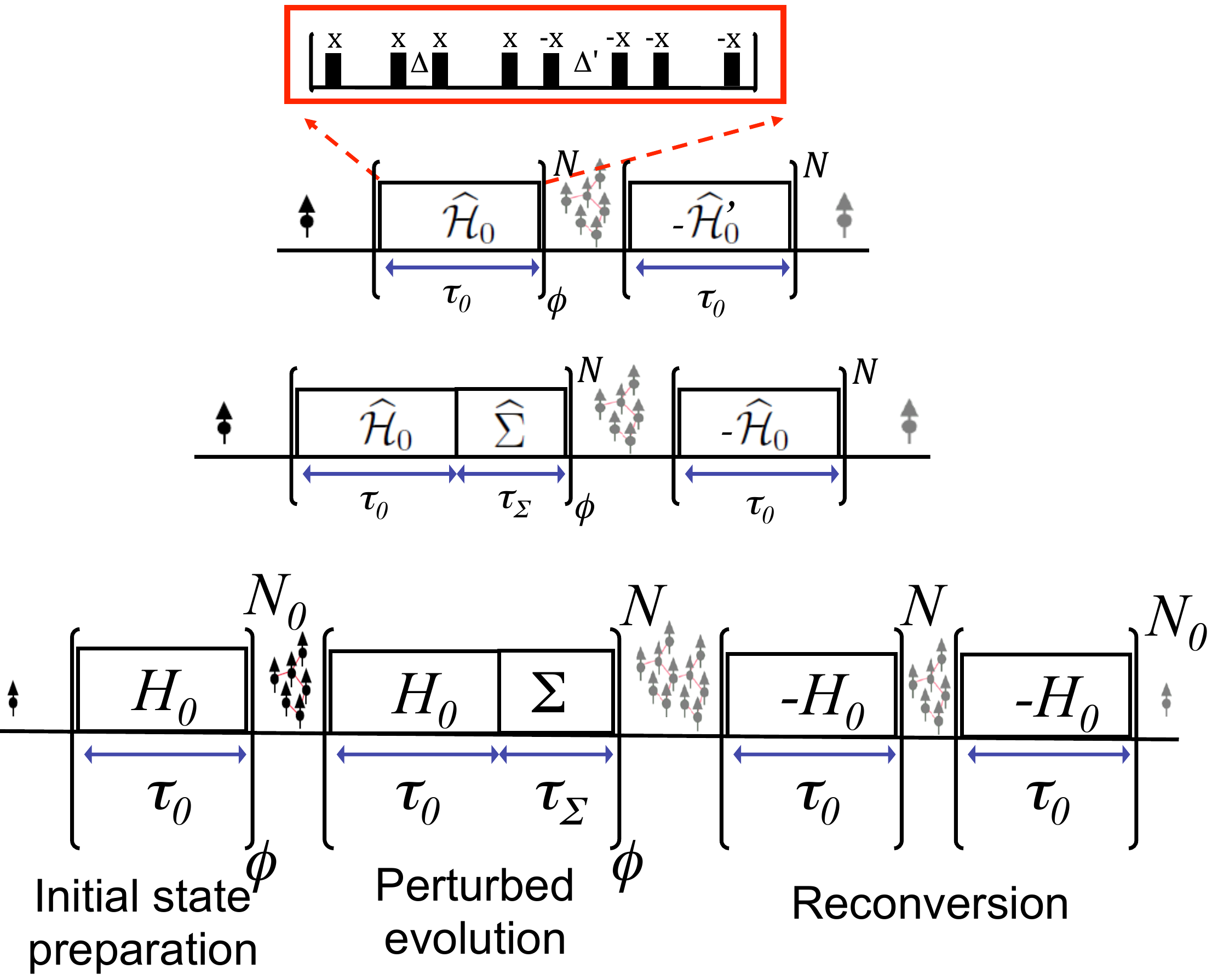}

\caption{(Color online) Sequence for generating a perturbed evolution. It is
achieved when $\tau_{\Sigma}\ne0$, where $\widehat{\Sigma}=\widehat{\mathcal{H}}_{dd}$
is the free evolution Hamiltonian.}

\label{Flo:NMRseqH0-perturbed} 
\end{figure}

The echo decay in Fig. \ref{fig:echodecay} depends of perturbation
$\widehat{\mathcal{H}}_{e}$. In order to study the sensitivity to
the perturbation strength, we introduced a perturbation $\widehat{\Sigma}$,
whose strength we can control experimentally and study the behavior
of the system as a function of the perturbation strength. We choose
the raw dipole-dipole coupling for this perturbation, $\widehat{\Sigma}=\mathcal{\widehat{H}}_{dd}$,
which has long range interactions with coupling strengths decaying
as $1/r^{3}$. We add this Hamiltonian to the ideal Hamiltonian $\widehat{\mathcal{H}}_{0}$
by concatenating short evolution periods under $\widehat{\mathcal{H}}_{dd}$
with evolution periods under $\widehat{\mathcal{H}}_{0}$. We label
the durations of the two time periods $\tau_{\Sigma}$ and $\tau_{0}$,
as shown in Fig. \ref{Flo:NMRseqH0-perturbed}.
 When the duration $\tau_{\mathrm{c}}=\tau_{0}+\tau_{\Sigma}$ of
each cycle is short compared to the inverse of the dipolar couplings
$d_{ij}$, the resulting evolution can be described by the effective
Hamiltonian
\begin{equation}
\widehat{\mathcal{H}}_{\mathrm{eff}}=(1-p)\widehat{\mathcal{H}}_{0}+p\widehat{\Sigma},\label{Heff-1}
\end{equation}
where the relative strength $p=\tau_{\Sigma}/\tau_{\mathrm{c}}$ of
the perturbation $\widehat{\Sigma}=\widehat{\mathcal{H}}_{dd}$ can
be controlled by adjusting the duration $\tau_{\Sigma}$. For the
quantum simulations, we compared the artificially perturbed evolution
of $\widehat{\mathcal{H}}_{\mathrm{eff}}$ with the $\widehat{\mathcal{H}}_{0}$
evolution with its intrinsic errors. While the intrinsic errors reduce
the signal or the overall fidelity, they do not cause localization
on the time scale of our experiments (see Fig. \ref{fig:Growth}).

\begin{figure}
\includegraphics[width=1\columnwidth]{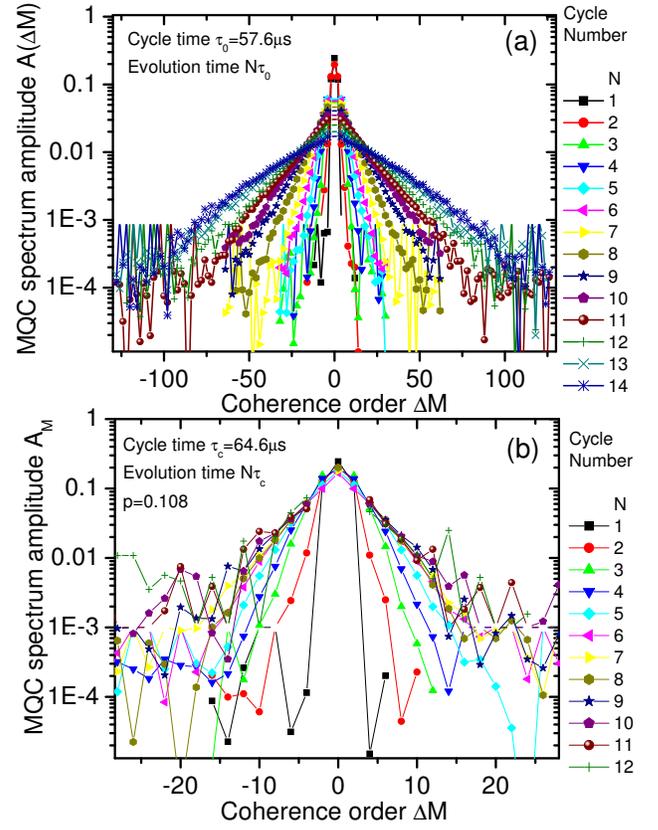}

\caption{\label{fig:Comparison of AM}(Color online) MQC spectrums for different
evolution times and perturbation strengths. Panel (a) shows the $A(\Delta M)$
spectrum for the unperturbed evolution ($p=0$) for different times.
Panel (b) shows its analogous but for $p=0.108$. Panel (b) shows
manifestation of the localization effects evidenced on the saturation
of the spreading of the MQC spectrum.}
\end{figure}

Starting from thermal equilibrium, now the state of the system at
the end of $N$ cycles is $\hat{\rho}^{\mathcal{H}_{\mathrm{eff}}}\left(N\tau_{\mathrm{c}}\right)=e^{-i\widehat{\mathcal{H}}_{\mathrm{eff}}N\tau_{\mathrm{c}}}\hat{\rho}_{0}e^{i\widehat{\mathcal{H}}_{\mathrm{eff}}N\tau_{\mathrm{c}}}.$
Taking into account now the complete sequence of evolutions given
by Fig. \ref{Flo:NMRseqH0-1}b, the experiment is thus a perturbed
forward evolution and an unperturbed backward evolution. The density
matrix at the end of the sequence is then $\sum_{M}\left[e^{i\mathcal{\widehat{H}}_{0}N\tau_{0}}\hat{\rho}_{M}^{\mathcal{H}_{\mathrm{eff}}}\left(N\tau_{\mathrm{c}}\right)e^{-i\mathcal{\widehat{H}}_{0}N\tau_{0}}\right]e^{iM\phi}$
as derived in \cite{alvarez_nmr_2010,alvarez_localization_2011}.
Thus the NMR echo signal, which is measured after the last backward
evolution $\exp\left\{ i\widehat{\mathcal{H}}_{0}N\tau_{0}\right\} $,
can be written as
\begin{multline}
\left\langle I_{z}\right\rangle \left(\phi,N\tau_{\mathrm{c}}\right)=\mbox{Tr}\left\{ \hat{I}_{z}\hat{\rho}_{f}\left(N\tau_{\mathrm{c}}+N\tau_{0}\right)\right\} \\
=\mathrm{Tr}\left\{ \hat{\rho}^{\mathcal{H}_{0}}\left(N\tau_{0}\right)\hat{\rho}_{\phi}^{\mathcal{H}_{\mathrm{eff}}}\left(N\tau_{\mathrm{c}}\right)\right\} .\label{eq:M-fidelities}
\end{multline}
In terms of the individual MQ coherences, this may be written as 
\begin{equation}
\left\langle I_{z}\right\rangle \left(\phi,N\tau_{\mathrm{c}}\right)=\sum_{\Delta M}\mbox{\ensuremath{e^{i\phi\Delta M}}Tr}\left\{ \hat{\rho}_{\Delta M}^{\mathcal{H}_{0}}\left(N\tau_{0}\right)\hat{\rho}_{\Delta M}^{\mathcal{H}_{\mathrm{eff}}}\left(N\tau_{\mathrm{c}}\right)\right\} 
\end{equation}
with the MQ coherence amplitudes $A(\Delta M)=\mbox{Tr}\left\{ \hat{\rho}_{\Delta M}^{\mathcal{H}_{0}}\left(N\tau_{0}\right)\hat{\rho}_{\Delta M}^{\mathcal{H}_{\mathrm{eff}}}\left(N\tau_{\mathrm{c}}\right)\right\} $.
For the ideal evolution ($p=0$), Eq. (\ref{eq:unperturbed_signal-1})
is recovered, where $A(\Delta M)$ correspond to the squared amplitudes
of the density operator elements $\hat{\rho}_{\Delta M}^{\mathcal{H}_{0}}\left(N\tau_{0}\right)$
with coherence order $\Delta M$. For the perturbed evolution, $(p\ne0)$,
they are reduced by the overlap of the actual density operator elements
$\hat{\rho}_{\Delta M}^{\mathcal{H}_{\mathrm{eff}}}\left(N\tau_{\mathrm{c}}\right)$
with the ideal ones. We extract these amplitudes by performing a Fourier
transformation with respect to $\phi$. Figure \ref{fig:Comparison of AM}
shows a comparison between the distributions $A(\Delta M)$ for different
evolution times for an unperturbed evolution (panel a) and a perturbed
evolution with $p=0.108$ (panel b). The main difference is that the
MQC spectrum of the perturbed evolution does not spread indefinitely
but its width reaches a limiting value \cite{alvarez_nmr_2010,alvarez_localization_2011}.

\begin{figure}
\includegraphics[width=1\columnwidth]{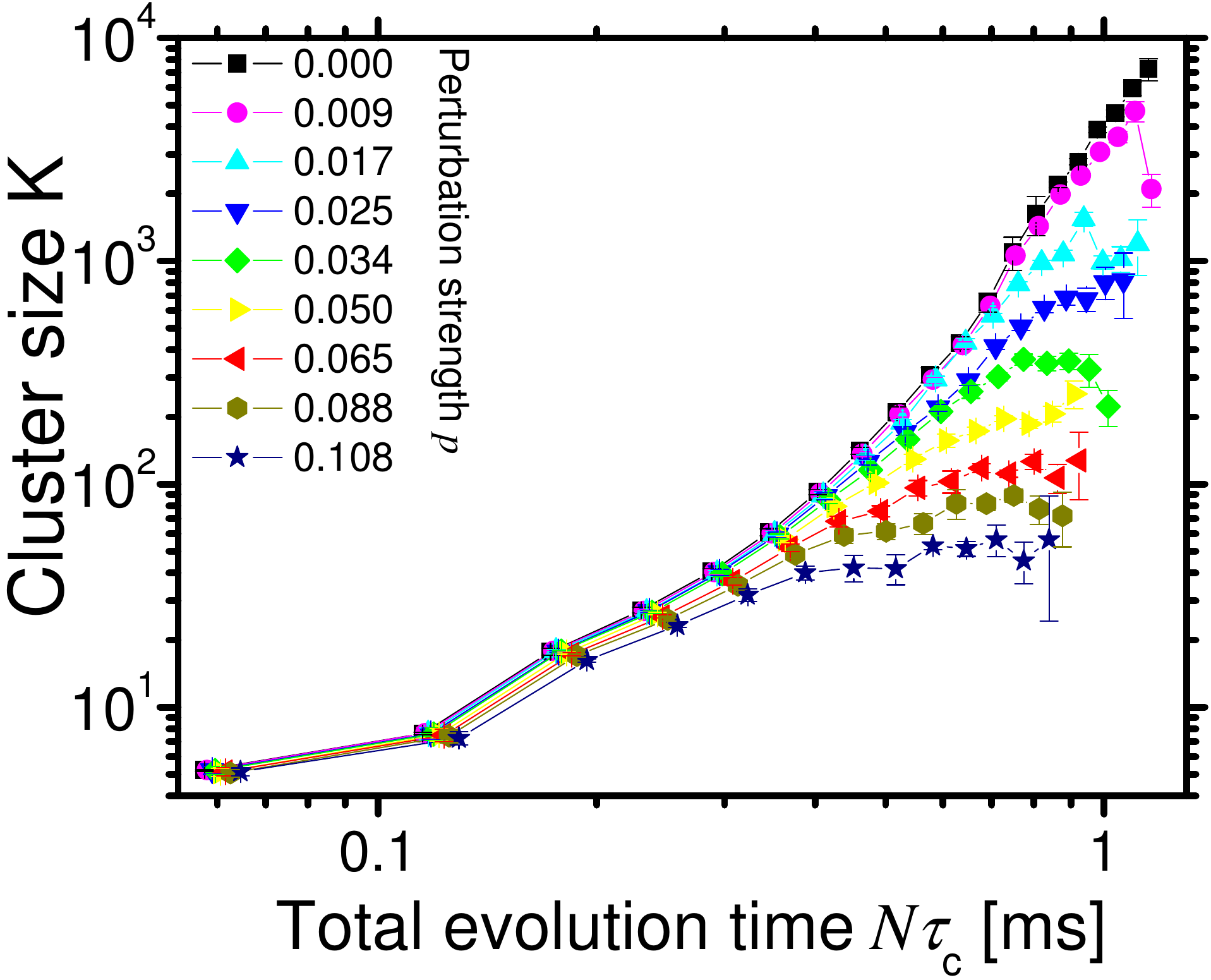}

\caption{\label{fig:cluster_sizesvsp}(Color online) Time evolution of the
cluster size $K$ for different perturbation strengths. The cluster
size is related to the volume occupied by the $K$ spins. By increasing
the perturbation, the localization-size is reduced. }
\end{figure}

As discussed in section \ref{sub:MeasureClusterSize}, we determine
the cluster size for different evolution times and perturbation strengths
from the width of the measured MQC distributions. Figure \ref{fig:cluster_sizesvsp}
shows the cluster size (the number of correlated spins) as a function
of the evolution time $N\tau_{c}$. The main difference of the perturbed
time evolutions (colored symbols in Fig. \ref{fig:cluster_sizesvsp})
compared to the unperturbed evolution (black squares) is that the
cluster size does not grow indefinitely \cite{alvarez_nmr_2010,alvarez_localization_2011},
but saturates. It remains unclear if the cluster growth for the weakest
perturbation $p=0.009$ also saturates. We consider this saturation
as evidence of localization due to the perturbation and the localization
size decreases with increasing the perturbation strength $p$. .

\subsection{Quantum dynamics from different initial cluster sizes}

\begin{figure}
\includegraphics[bb=0bp 570bp 605bp 770bp,clip,width=1\columnwidth]{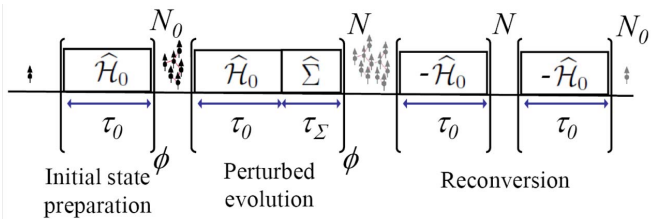}

\caption{(Color online) Sequence for preparing different initial clusters sizes
by controlling $N_{0}$ and subsequently evolving them in the presence
of a perturbation. }

\label{Flo:NMRseqdiffinitialstates} 
\end{figure}

We have shown that time evolution of the cluster size under perturbations
reaches a dynamical equilibrium state \cite{alvarez_nmr_2010,alvarez_localization_2011},
i.e. for a given perturbation strength, the size of the spin clusters
tends toward the same limiting value, independent of the initial condition.
In order to show this, we prepared a series of initial conditions
corresponding to different clusters sizes. Figure \ref{Flo:NMRseqdiffinitialstates}
shows the corresponding pulse sequence: The initial state preparation,
consisting of an evolution of duration $N_{0}\tau_{0}$ under the
unperturbed Hamiltonian $\widehat{\mathcal{H}}_{0}$, generates clusters
of size $K_{0}$.

\begin{figure}
\includegraphics[width=1\columnwidth]{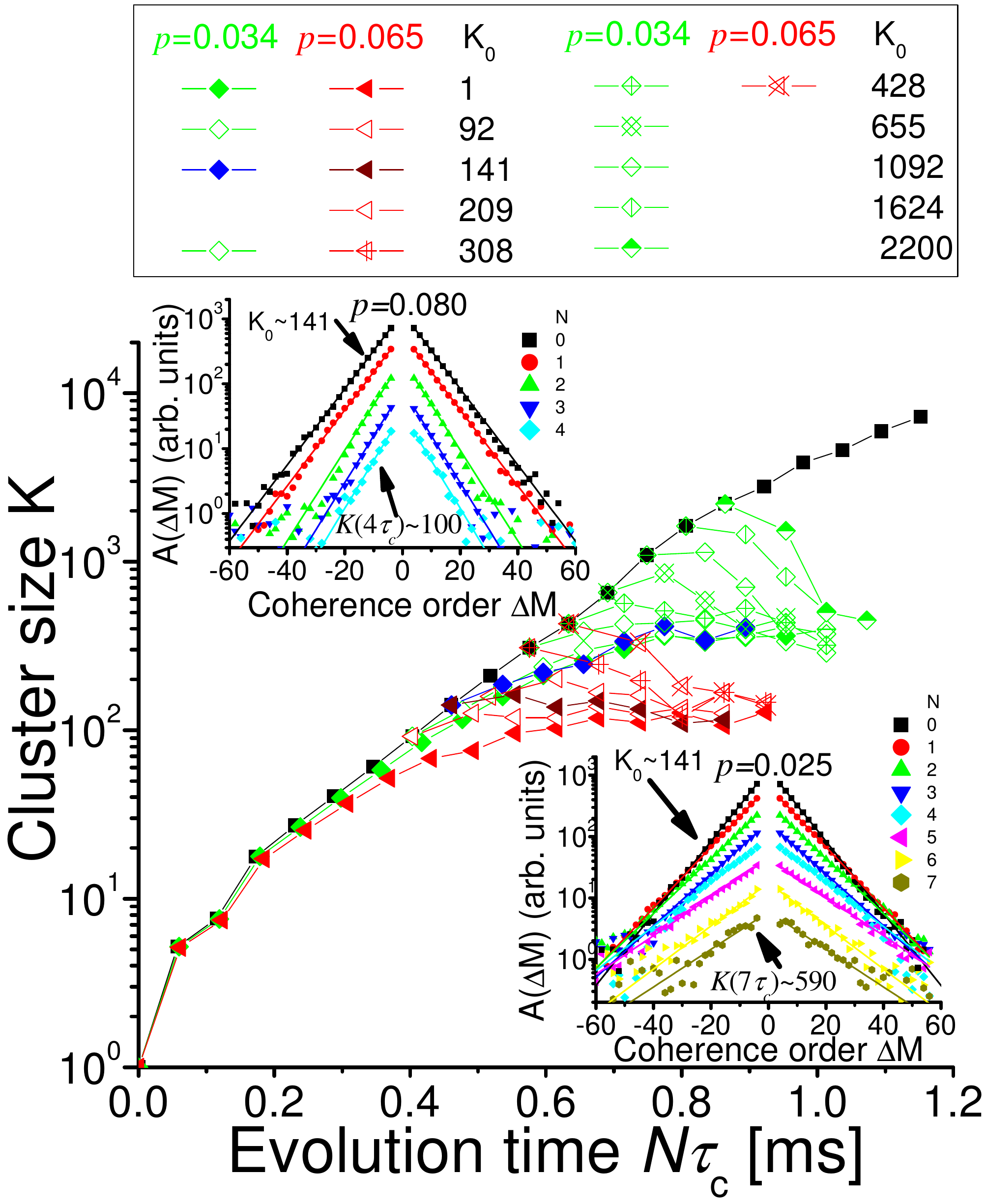}

\caption{\label{fig:dyneq-ams}(Color online) Time evolution of the cluster-size
of correlated spins starting from different initial sates. The experimental
data is shown for two different perturbation strengths given in the
legend. The solid black squares, red triangles and green rhombuses
are evolutions from an uncorrelated initial state. Empty symbols start
from an initial state with $K_{0}$ correlated spins (see legend).
The solid blue rhombuses and brown triangles starts from an initial
$K_{0}=141$. The insets show the MQC spectrum starting from $K_{0}=141$
as a functions of time for the two perturbation strengths.}
\end{figure}

 During the subsequent perturbed evolution of duration $N\tau_{\mathrm{c}}$,
these initial clusters evolve and Eq. (\ref{eq:M-fidelities}) becomes

\begin{multline}
\left\langle I_{z}\right\rangle \left(\phi,N_{0}\tau_{0}+N\tau_{\mathrm{c}}\right)=\\
\mbox{Tr}\left\{ \hat{\rho}^{\mathcal{H}_{0}}\left(N\tau_{0},N_{0}\tau_{0}\right)\hat{\rho}_{\phi}^{\mathcal{H}_{\mathrm{eff}}}\left(N\tau_{c},N_{0}\tau_{0}\right)\right\} .\label{eq:Izdiffinitialstates}
\end{multline}
This method allows us to study the growth of the clusters by starting
from different sizes $K_{0}=K(N_{0}\tau_{0})$ and following the evolution
as a function of time and perturbation strength. Based on Eq. (\ref{eq:Izdiffinitialstates}),
we determined the MQC spectra $A(\Delta M)$ for different evolution
times. The insets of Fig. \ref{fig:dyneq-ams}, shows two examples
of them. From such curves, we determined the evolution of the cluster
size shown in Fig. \ref{fig:dyneq-ams}. The figure shows evolution
of the the cluster size for two perturbation strengths, starting from
different initial sizes. The dynamical equilibrium is clearly manifested
in the figure. The two insets show the $A(\Delta M)$ spectrums starting
from $K_{0}=141$, for different evolution times. We can see that
if $K_{0}$ is lower than the localization size, the MQC spectrum
spreads until it localizes (manifested by the parallel slopes), however
if $K_{0}$ is larger than the localization size, it shrink until
saturation. We found that the localization size vs. the perturbation
strength is roughly proportional to $1/p^{2}$ \cite{alvarez_nmr_2010,alvarez_localization_2011}.
During our experiment, the magnetization is uniform throughout the
sample, so the process does not lead to a spatial redistribution of
magnetization. Note however that here we measure the cluster size
of correlated spins that is associated with a coherent length. Therefore,
this technique allows to investigate the localization size, even when
the density profile of the excited cluster size would exceed the localization
length. 

\section{Conclusions}

As a step toward the understanding of the quantum evolution of large
quantum systems, we have studied the spreading of information in a
system of nuclear spins. Decoherence has long been recognized to limit
the time for which quantum information can be used. Spatial disorder
also limits the distance over which quantum information can be transferred.
We have studied the role of a disordered dipolar interaction Hamiltonian
and shown that for larger values of the perturbation, the coherence
length of the cluster size reaches a limit value. Even though we do
not measure directly the spatial extent of the cluster size, one might
speculate that the spatial extent and the number of correlated spins
are related with Volume$\sim K$. A connection to Anderson localization
of spins with dipolar $1/r^{3}$ interactions could thus be explored
\cite{Anderson1958,anderson_local_1978,Pomeransky2004,Burrell2007,Keating2007,Allcock2009}.
We also note that for lower values of the perturbation, the size of
the cluster grows faster than expected for a diffusive model. We will
investigate possible connections to Levy flights and Levy walks induced
by the long range dipole-dipole couplings of our Hamiltonians.We developed
a method that allows one to quantify the time evolution of the cluster
size of correlated spins starting with single qubits. As we have shown,
the information can spread to clusters of several thousand qubits.
We have observed that the combination of an information spreading
Hamiltonian and a perturbation to it results in a quantum state that
becomes localized. The localization size decreases with increasing
strength of the perturbation and the resulting size appears to be
determined by a dynamic equilibrium \cite{alvarez_nmr_2010,alvarez_localization_2011},
a feature which might be adapted to other communities studying Anderson
localization. 

The results presented here provide information about the spatial bounds
for transferring quantum information in large spin networks and indicate
how precise manipulations of large quantum systems have to be. The
sample used in this study is also an interesting system for studying
fundamental aspects of Anderson localization in the presence of long
range interactions.
\begin{acknowledgments}
This work was supported by the DFG through Su 192/24-1. G.A.A. acknowledges
financial support from the Alexander von Humboldt Foundation and from
the European Commission under the Marie Curie Intra-European Fellowship.
\end{acknowledgments}
\bibliographystyle{apsrev4-1}
\bibliography{localization_long,bibannalen1,bibannalen2}

\end{document}